\documentclass{icrc2009}

\usepackage{graphicx}   
\usepackage{caption}    
\usepackage[font=footnotesize]{subfig} 
\usepackage{fixltx2e}
\usepackage{url}

\newcommand{\shorttitle}[1]%
{\markboth{Proceedings of the 31\MakeLowercase{$^{st}$} ICRC, {\L}\'{o}d\'{z} 2009}{#1} }
\newcommand{\etal}{\MakeLowercase{\textit{et al. }}} 

\begin{document}
\title{Simultaneous Observations of Flaring Gamma-ray Blazar 3C 66A with {\em Fermi}-LAT and VERITAS }

\author{\IEEEauthorblockN{Luis C. Reyes\IEEEauthorrefmark{1} on behalf of the {\em Fermi}-LAT and VERITAS\IEEEauthorrefmark{2} Collaborations}
                            \\
\IEEEauthorblockA{\IEEEauthorrefmark{1}Kavli Institute for Cosmological Physics (KICP) at The University of Chicago, Chicago, IL 60605, USA}
\IEEEauthorblockA{\IEEEauthorrefmark{2}see R.A. Ong et al. (these proceedings) or http://veritas.sao.arizona.edu/conferences/authors?icrc2009
}}

\shorttitle{L. C. Reyes \etal MW Observations of 3C66A}
\maketitle

\begin{abstract}
 The intermediate-frequency-peaked BL Lac object 3C 66A was detected in a flaring state by the {\em Fermi}-LAT and VERITAS observatories in October 2008. These data and follow-up observations at other wavelengths create a rich sample of light curves and a constraining spectral energy distribution (SED). This is the first time that simultaneous observations at GeV and TeV energies were obtained for a flaring blazar. Results from these joint {\em Fermi}-LAT and VERITAS observations are presented in this paper.
  \end{abstract}

 
\section{Introduction}


Due to its significant optical and X-ray variability, 3C 66A was classified as a BL Lac object by Maccagni et al. \cite{key:Maccagni}, and given the location of its synchrotron peak (between $10^{15}$ and $10^{16}$ Hz), 3C 66A can be further sub-classified as an intermediate-frequency peaked BL Lac (IBL). BL Lacs are known for having very weak (if any) detectable emission lines, which makes the redshift determination quite difficult. The redshift of 3C 66A was reported as $z=0.44$ by Miller et al. \cite{key:Miller} and independently by Lanzetta et al. \cite{key:Lanzetta}. However, both measurements rely on the measurement of a single line, and as pointed out by Bramel  et al. \cite{key:Bramel}, the redshift of this object is still quite uncertain.  


Similarly to other blazars, the spectral energy distribution (SED) of 3C 66A is known to have two pronounced peaks (see \cite{key:Joshi_Boettcher} for SED details), which suggests at least two different physical emission processes at play. The first peak (extending from radio to soft X-ray frequencies) is  likely due to polarized synchrotron emission from high energy electrons, while different emission models (briefly introduced below) have been proposed to explain the second energy peak, which extends up to gamma-ray energies. It should be noted that the high-energy emission component dominates the bolometric luminosity of this source, a powerful demonstration of the extreme nature of this type of object.

The models that have been proposed to explain gamma-ray emission in blazars can be roughly categorized into leptonic and hadronic mechanisms, depending on whether the accelerated particles responsible for the gamma-ray emission are primarily electrons/positrons or protons. In leptonic models, high-energy electrons/positrons produce gamma rays via inverse Compton scattering of low-energy photons.  In synchrotron self-Compton (SSC) models, the same population of electrons responsible for the observed gamma rays generates the low-energy photon field through synchrotron emission. In external Compton (EC) models the low-energy photons originate outside the jet. Possible sources of external photons include: accretion disk photons radiated directly into the jet, accretion disk photons scattered by emission-line clouds or dust into the jet, or synchrotron radiation re-scattered back into the jet by broad-line emission clouds.  In hadronic models gamma rays are produced by high-energy protons, either via proton synchrotron radiation, or via secondary  (see \cite{key:Boettcher} and references therein for a review of blazar gamma-ray emission processes). 

One of the main obstacles in the broadband study of gamma-ray blazars is the  lack of simultaneity, or 
at least contemporaneity, of the data at the various wavelengths.  At high energies the situation has been even more difficult due to the lack of objects that can be detected by MeV/GeV and TeV observatories in comparable time scales. Indeed, until recently most of our  knowledge of blazars at gamma-ray energies was obtained from observations performed in two disjoint energy regimes:  i) the high energy (HE) range (20 MeV$<E<$ 10 GeV), studied in the 1990s by EGRET \cite{key:EGRET}, and ii) the very-high-energy (VHE) regime (E $>$ 100 GeV) observed by ground-based instruments \cite{key:Weekes_TevReview}. Remarkably, most EGRET blazars have not been detected by TeV telescopes; even those that are nearby and bright \cite{key:Fegan}. Furthermore, blazars detected by EGRET at MeV/GeV energies are predominantly FSRQs, while blazars TeV blazars are  BL Lacs. It is important to understand these observational differences since they are likely related to the physics of the AGN \cite{key:Cavalieri}, or to the evolution of blazars over cosmic time \cite{key:Boettcher_Dermer}. 

Fortunately, the new generation of gamma-ray instruments (AGILE, {\em Fermi}, HESS, MAGIC and VERITAS) is closing the gap between both energy regimes due to their improved sensitivities, leading us towards a deeper and more complete characterization of blazars as a high-energy source and as a population.  An example of the successful synergy of space-borne and ground-based observatories  was provided by the joint observations of 3C 66A by {\em Fermi} and VERITAS during its strong flare of October 2008.  The strong flare was originally reported by VERITAS \cite{key:VERITAS_ATEL}, and soon after,  correlated variability was detected in the {\em Fermi}-LAT data \cite{key:Fermi_ATEL}. Follow-up observations were obtained at radio, optical and X-ray wavelengths in order to measure the flux and spectral variability of the source across the electromagnetic spectrum and to obtain a quasi-simultaneous SED. In this contribution we report preliminary results of this campaign, the resulting broadband spectrum and our first attempt to model this very constraining SED.

\begin{figure}[htbp]
\centering
\includegraphics[width=0.45\textwidth]{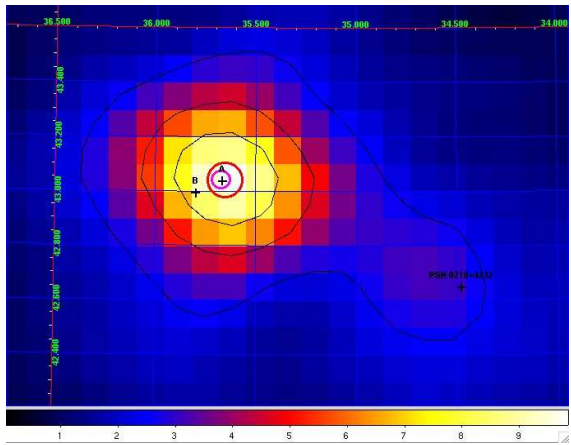}
\caption{{\footnotesize Smoothed count map of the 3C 66A region as seen by {\em Fermi} between Sep 1 and Dec 31 2008 with E$>$100 MeV. The color bar has units of counts per pixel and the pixel dimensions are  0.1 x 0.1 degrees. The locations of 3C 66A and 3C 66B are shown as crosses and labeled ``A'' and ``B'' respectively. The location of PSR 0218+4232 is also indicated.   The magenta circle represents the VERITAS localization of the VHE source (2h 22m 41.6s $\pm $1.7s (stat) $\pm $6.0s (sys) ; $43^{o}$ 02' 35.5" $\pm$ 21" (stat) $\pm$1'30" (sys)) as reported in Acciari et al. \cite{key:veritas_paper}.  The red circle represents the 95\% error radius of the {\em Fermi} localization (RA, DEC) = (35.6541,  43.0402) and includes statistical error  $0.042^{o}$ and a conservative systematic error of $0.04^{o}$ as discussed in Abdo et al. \cite{key:brigh_source_list}. Although the {\em Fermi} localization clearly favors blazar 3C 66A as the gamma-ray source counterpart, some small contribution from radio galaxy 3C 66B \cite{key:tavecchio} cannot be excluded. A more detailed analysis is ongoing to determine the likelihood of such a scenario.}}
\label{default}
\end{figure}

\section{Observations and Data Analysis}

{\bf VERITAS:} The Very Energetic Radiation Imaging Telescope Array System (VERITAS) is an array of four 12m diameter  imaging atmospheric Cherenkov telescopes (IACTs) in southern Arizona, U.S.A. \cite{key:Acciari08}.  VHE emission from 3C 66A was detected by  VERITAS  from 33 hours of observations between September and November 2008. A complete description of the VERITAS analysis and results is available in Acciari et al. \cite{key:veritas_paper}. It suffices to summarize here that a strong flare with day-scale variability was observed in October 2008 (as illustrated in the bottom panel of fig. 2) and that the observed spectrum is very steep (as shown in fig. 3), yielding an index of 4.1 $\pm$ 0.4 (stat) $\pm$ 0.6 (sys) when fitted to a power law.

{\bf {\em Fermi}-LAT: } The data from the Large Area Telescope (LAT; \cite{key:Atwood_LAT}) have been analyzed by using ScienceTools v9r11. Only {\em class-3} events (also called {\em diffuse})  were selected for this analysis because of their negligible background contamination and very good angular reconstruction. In order to avoid Earth's albedo photons, a zenith angle $<$105 degrees cut in instrument coordinates was used. The diffuse emission from the Galaxy was modeled using a GALPROP model \cite{key:GALPROP} which has been refined with {\em Fermi}-LAT data taken during the first 3 months of operation. The extragalactic diffuse and residual instrumental backgrounds have been modeled as an isotropic component and included in the fit (see \cite{key:brigh_source_list} for a complete discussion of the treatment of these backgrounds).  The data were analyzed with an unbinned maximum likelihood technique \cite{key:Mattox} using the likelihood analysis software developed by the LAT team. Although 3C 66A was detected by EGRET as source 3EG J0222+4253 \cite{key:3rdEgret}, Kuiper et al.  \cite{key:Kuiper} showed with detailed spatial and timing analyses that this EGRET source actually consists of the superposition of 3C 66A and nearby millisecond pulsar PSR J0218+4232. This interpretation of the EGRET data has been verified by {\em Fermi}-LAT, whose improved angular resolution permits the clear separation of the two sources as shown in Fig. 1. Furthermore, gamma-ray pulsations from PSR J0218+4232 have been detected with high confidence by the {\em Fermi} team and will be reported elsewhere \cite{key:ms_pulsars}.  More importantly for this analysis, the clear separation between the pulsar and the blazar allows us to account for each source independently in the maximum likelihood analysis and thus, permits an accurate determination of the spectrum and localization of each source.

{\bf Chandra: }3C66A was observed by the Chandra Observatory on October 6th 2008 for a
total of 37.6~ksec with the Advanced CCD Imaging Spectrometer (ACIS), covering
the energy band between 0.3 and 10~keV. The source was observed in the continuous clocking (CC) mode to avoid pile-up effects. The standard data processed files provided by the
Chandra X-ray Center (processed with software ver. 7.6.11.9 and calibration
files (CALDB) ver. 3.5.0) have been used. 

\begin{figure}[htbp]
\centering
\includegraphics[width=0.5\textwidth]{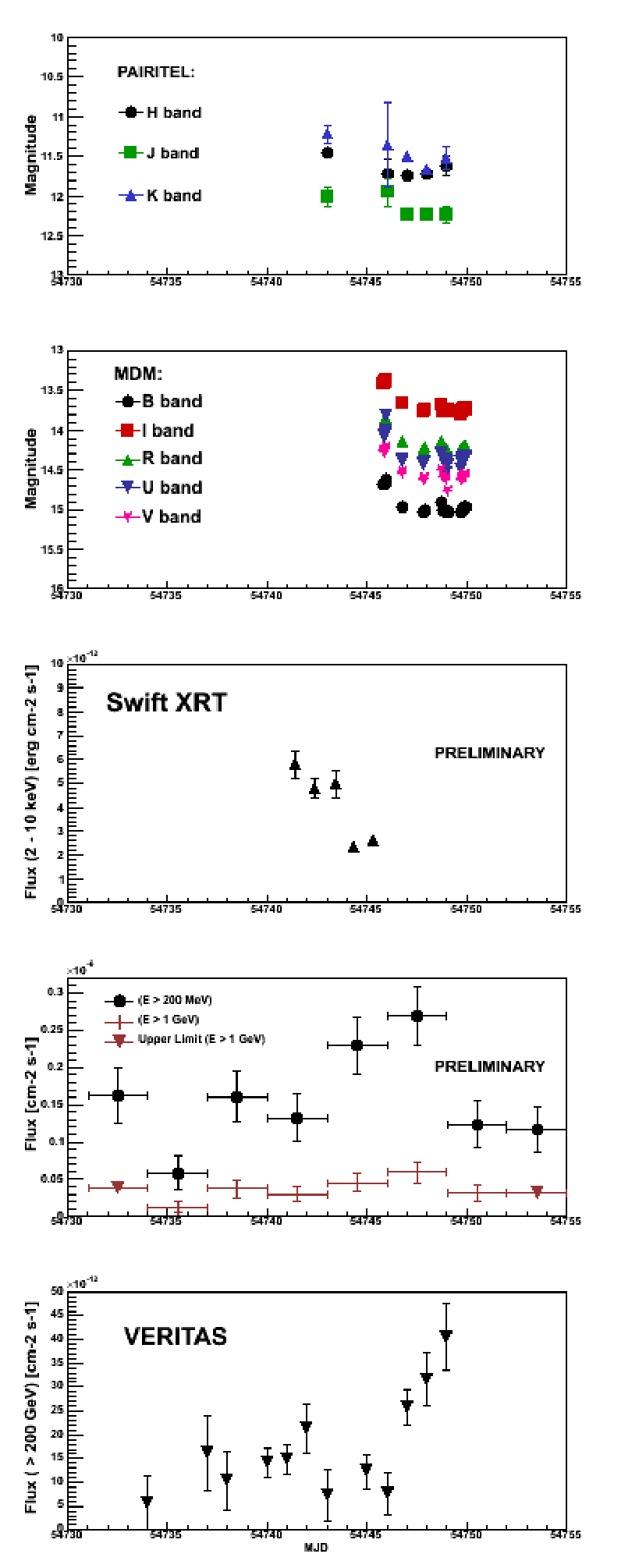}
\caption{{\footnotesize 3C 66A light curves from (top to bottom): PAIRITEL, MDM, Swift XRT, {\em Fermi} and VERITAS. The {\em Fermi} panel includes light curves for E $>$ 200 MeV and E $>$ 1 GeV using  time bins with width equal to 3 days.  It should be noted that the flare observed by VERITAS coincides with the period of time when {\em Fermi} detected the source at high energies ($>$ 1 GeV). }}
\label{default}
\end{figure}

{\bf Swift XRT and UVOT:}  Following the VERITAS detection of VHE emission from 3C 66A, Target of Opportunity (ToO) observations of 3C66A with Swift were obtained for a total duration of $\sim10$ksec. Swift comprises an UV-Optical telescope UVOT, an X-ray telescope (XRT) and a Burst Alert Telescope \cite{key:swift}. Data reduction and calibration of the XRT data were performed with {\em HEASoft} v6.5 standard tools. All XRT data presented here were taken in Photon Counting (PC) model with negligible pile-up effects. The X-ray spectra of each observation were fit with an absorbed power law using a fixed Galactic column density from Dickey \& Lockman \cite{key:Dickey_Lockman}, which gave good $\chi^{2}$ values for all observations. UVOT observations were taken over six  bands (V, B, U, W1, M2 and W2) that were calibrated using standard techniques \cite{key:Poole, key:Li} and  redshift corrected using a nominal value of $z=0.44$.

{\bf MDM Optical Observations:} 3C 66A was observed at the 1.3m telescope of the 
MDM Observatory during the nights of Oct. 6 -- 10,
2008 UT. A total of 290 science frames in U, B,
V, R, and I bands (58 each) were taken throughout
the entire visibility period (approx. 4:30 -- 10:00
UT) during each night. 

{\bf PAIRITEL Observations:}  Near-infrared observations in the  {\em J}, {\em H} and {\em $K_{s}$} were obtained with the 1.3m Peters Automated Infrared Imaging Telescope located at the Fred Lawrence Whipple Observatory.  As a fully robotic observatory, PAIRITEL relies on a  software system that manages  the telescope and reduces data as it is gathered.  See Bloom et al. \cite{key:Bloom} for a full description of the data reduction procedure.

Additional multiwavelength observations at radio, optical and X-ray wavelengths  will be included in a future paper.

\section{Discussion}
 
The measured SED of 3C66A in October 2008 is presented in Fig. 3. We have modeled the SED using the code of B\"ottcher \& Chiang \cite{key:Boettcher_Chiang}. In this
model, a power-law distribution of ultrarelativistic electrons
and/or pairs with lower and upper energy cutoffs at $\gamma_{\rm min}$
and $\gamma_{\rm max}$, respectively, and power-law index $q$
is injected into a spherical region of co-moving radius $R_B$. 
The injection rate is normalized to an injection luminosity
$L_e$, which is a free input parameter of the model. A 
temporary equilibrium between particle injection, radiative
cooling due to synchrotron and Compton losses, and particle
escape on a time scale $t_{\rm esc} \equiv \eta_{\rm esc}
\, R_B/c$ is assumed. Both the internal synchrotron photon
field (SSC) and external photon 
sources (EC) are considered as targets 
for Compton scattering. The emission region is moving with
a bulk Lorentz factor $\Gamma$ along the jet. To reduce the
number of free parameters, we assume that the jet is oriented
with respect to the line of sight at the superluminal angle
so that the Doppler factor is equal to $D = \left( \Gamma \, [1 - \beta
\, \cos\theta_{\rm obs} ] \right)^{-1}$ where $\theta_{\rm obs}$ is the angle of the jet with respect to the line of sight.  We use the extragalactic-background-light model of Franceschini et al. \cite{key:Franceschini} and the adopted value of $z = 0.444$ for our model, but keep in mind the caveat that the results might be significantly different if the source is at a very different redshift.

\begin{figure}[htbp]
\centering
\includegraphics[width=0.45\textwidth]{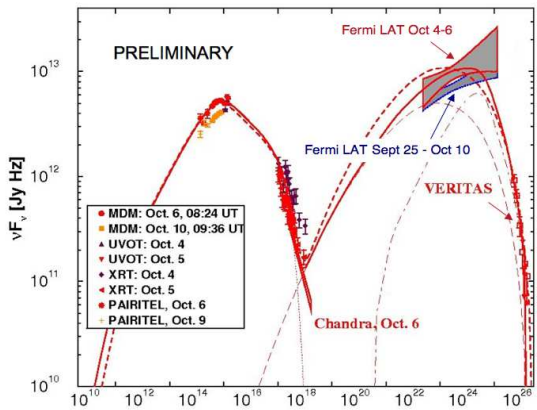}
\caption{{\footnotesize  SED of 3C 66A including {\em Fermi}-LAT and VERITAS data. The {\em Fermi} spectrum was calculated for the short time interval  Oct 4 - 6  (MJD 54743-54745) where most of the multiwavelength observations took place and for a longer time interval corresponding to the VERITAS {\em dark period}  where VHE emission from 3C 66A was detected (MJD 54734 - 54749). The {\em Fermi} flux from Oct 4 - 6  is higher F(E$>$100MeV) = $4.9\pm0.9$ x 10$^{-7}$cm$^{-2}$s$^{-1}$ than the {\em dark period} flux of 3.6 $\pm$ 0.4 x 10$^{-7}$cm$^{-2}$s$^{-1}$. In both cases the {\em Fermi} spectrum is quite hard and can be described by a power law with index 1.84 $\pm$ 0.11 and 1.87 $\pm$ 0.06 in the Oct 4-6 and {\em dark period} intervals respectively,  thus, no spectral evolution is evident in the {\em Fermi} data. Furthermore, it should be noted that there are not enough photons at high-energies in the {\em Fermi} data (6 photons with E $>$ 10 GeV for the Oct 4 - 6 period) to confirm the spectral steepening around 10 GeV indicated by the model. }}
\label{default}
\end{figure}

  \begin{table}[!h]
  \label{table_wide}
  \centering
  \begin{tabular}{|c|c|c|}
  \hline
   Parameter & Symbol & Value \\
   \hline 
$e^{-}$ Injection Luminosity	& $L_e$			& $1.26\times10^{45}$~erg~s$^{-1}$ \\
Low-energy cut-off 	& $\gamma_{\rm min}$ 	& $5\times10^3$ \\
High-energy cut-off & $\gamma_{\rm max}$ 	& $8\times10^4$ \\
$e^{-}$ injection index 	& $q$			& $2.5$ \\
Plasmoid radius 		& $R_B$			& $2\times10^{16}$~cm \\
Bulk Lorentz factor $\Gamma$	& $\Gamma$		& $30$ \\
Observing Angle                                                      &  $\theta_{\rm obs}$  & 1.9 deg \\
Co-moving magnetic field	& $B$			& $0.3$~G \\
$e^{-}$ escape time scale & $\eta_{\rm esc}$	& $100$ \\
Ext. radiation energy density & $u_{\rm ext}$	& $6.6\times10^{-6}$~erg~cm$^{-3}$ \\
Ext. radiation peak frequency & $\nu_{\rm ext}$	& $1.5\times10^{14}$~Hz \\
\hline
  \end{tabular}
  \caption{{\footnotesize EC+SSC model fit parameters}}
  \end{table}

Most TeV blazars known to date are high-frequency peaked BL Lacs (HBLs), whose SEDs can often be fit satisfactorily with pure SSC models. Therefore,
we first attempted a pure SSC model to fit the near-IR
through VHE gamma-ray SED. However, we found that such
a model (dashed line in Fig. 3) grossly overproduces the gamma-ray flux at GeV energies and requires a magnetic field far below the value expected from equipartition.  A better fit to the entire SED 
could be achieved by including an additional external radiation
field, modeled as a thermal radiation field with a
peak frequency at $\nu_{\rm ext} = 1.5 \times 10^{14}$~Hz.
and a radiation density of $u_{\rm ext} = 6.6 \times 10^{-6}$
erg~cm$^{-3}$. Nevertheless, the EC+SSC model (solid line in Fig. 3) still requires a sub-equipartition magnetic field $(u_{B}/u_{e}\sim0.1)$ (the full set of parameters used for the fit
shown in Fig. 3 is listed in table 1). These
 preliminary results agree with the expectation that for IBLs the higher ambient  luminosity at optical wavelengths plays a fundamental role in providing a seed population of soft photons for inverse Compton scattering. 

\section{Conclusion and Future Plans}

Multiwavelength observations of 3C 66A were carried out following the gamma-ray outburst detected by the VERITAS and {\em Fermi} observatories in October 2008. This marks the first occasion that a gamma-ray flare is detected by GeV and TeV instruments in comparable time scales. The SED constructed with these and other contemporaneous multiwavelength observations constitutes a fertile ground for the modeling of high-energy emission in gamma-ray blazars.  A joint paper to further address this and other topics is in preparation by the {\em Fermi}-LAT and VERITAS collaborations. 

\section*{Acknowledgments}

{\footnotesize The {\em Fermi}-LAT collaboration acknowledges generous ongoing support from a number  of agencies and institutes that have supported both the development and the operation of the 
LAT as well as scientific data analysis. These include the National Aeronautics and Space 
Administration and the Department of Energy in the United States, the Commissariat `a 
l�Energie Atomique and the Centre National de la Recherche Scientifique / Institut National 
de Physique Nucleaire et de Physique des Particules in France, the Agenzia Spaziale Italiana 
and the Istituto Nazionale di Fisica Nucleare in Italy, the Ministry of Education, Culture, 
Sports, Science and Technology (MEXT), High Energy Accelerator Research Organization 
(KEK) and Japan Aerospace Exploration Agency (JAXA) in Japan, and the K. A. Wallenberg Foundation, the Swedish Research Council and the Swedish National Space Board in 
Sweden.  The VERITAS collaboration acknowledges the support from the US Department of Energy, the US National Science Foundation, and the Smithsonian Institution, by NSERC in Canada, by Science Foundation Ireland, and by STFC in the UK. The VERITAS collaboration also acknowledges the excellent work of the technical support staff at the FLWO and the collaborating institutions in the construction and operation of the instrument. PAIRITEL is operated by the Smithsonian Astrophysical Observatory (SAO) and was made possible by a grant from the Harvard University Milton Fund, a camera loan from the University of Virginia, and continued support of the SAO and UC Berkeley. The PAIRITEL project is further supported by NASA/Swift Guest Investigator grant NNG06GH50G. L. Reyes acknowledges the support by the Kavli Institute for Cosmological Physics at the University of Chicago through grants NSF PHY-0114422 and NSF PHY-0551142 and an endowment from the Kavli Foundation and its founder Fred Kavli.

\end{document}